\documentclass[proceedings]{JHEP3} 

\PrHEP{AHEP2003}

\def\np#1#2#3{           {\sl Nucl. Phys. }{\bf #1}, #2 (#3)}
\def\pl#1#2#3{           {\sl Phys. Lett. }{\bf #1}, #2 (#3)}
\def\pr#1#2#3{           {\sl Phys. Rev. }{\bf #1}, #2 (#3)}

\def\prl#1#2#3{          {\sl Phys. Rev. Lett. }{\bf #1}, #2 (#3)}
\def\jetp#1#2#3{         {\sl JETP Lett. }{\bf #1}, #2 (#3)}

\def\sc#1#2#3{         {\sl Science }{\bf #1}, #2 (#3)}

\def\mpl #1#2#3{         {\sl Mod.~Phys.~Lett. }{\bf #1}, #2 (#3)}
\def\etal{\hbox{\it et al.}}
\def\beq{\begin{equation}}
\def\eeq{\end{equation}}
\def\bc{\begin{center}}
\def\ec{\end{center}}
\def\matriz#1{\left( \begin{array}{ccc} #1 \end{array} \right)}

\def \lsim{\mathrel{\vcenter
     {\hbox{$<$}\nointerlineskip\hbox{$\sim$}}}}
\def \gsim{\mathrel{\vcenter
     {\hbox{$>$}\nointerlineskip\hbox{$\sim$}}}}

\conference{International Workshop on Astroparticle and High Energy Physics}

\title{Neutrino masses and GUT baryogenesis}

\author{\speaker{Juan A. Lopez-Perez}, Nuria Rius\\
        IFIC Valencia\\
        E-mail: \email{juanlope@ific.uv.es}, \email{nuria@ific.uv.es}}

\abstract{We reconsider the GUT-baryogenesis mechanism for generating the 
baryon asymmetry of the Universe. The baryon asymmetry is produced
by the out of equilibrium decay of coloured Higgs bosons at the GUT scale,
conserving B-L.
If neutrinos are Majorana particles, lepton number violating interactions 
erase the lepton number excess, but part of the baryon asymmetry may be 
preserved, provided those interactions are not in thermal equilibrium 
when the sphaleron processes become effective, at $T \sim 10^{12}~ GeV$.
We analyse whether this mechanism for baryogenesis is feasible in a 
variety of GUT models of fermion masses proposed in the literature, 
based on horizontal symmetries.}

\begin{document}

{\flushright IFIC/04-14,FTUV/04-0404}

  \section{Introduction}

	In the GUT-baryogenesis mechanism \cite{gut-baryo} 
the baryon asymmetry is generated during the out of 
equilibrium decay of coloured Higgs bosons, 
because those decays violate $B$ and $L$.
The problem is that in $SU(5)$ grand unification, 
and in any grand unified model 
containing $U(1)_{B-L}$ as a subgroup, the dominant Higgs decays
conserve $(B-L)$. 
Below $10^{12}~ GeV$ the sphaleron configurations \cite{sph} are 
in thermal equilibrium \cite{spheq} and violate also $B$ and $L$, but 
conserve $(B-L)$, relating these quantities as
\beq
B=c~(B-L)=\frac{c}{c-1}~L
\eeq
where c is a constant. As a result the baryon asymmetry generated 
in the Higgs decays is erased, since $(B-L)=0$. 
    However, recently it has been proposed a modified GUT baryogenesis 
mechanism which may produce the observed baryon asymmetry, provided neutrinos 
are Majorana particles \cite{resurrection}.

 It is now confirmed that 
the  deficits in atmospheric muon neutrinos \cite{fukuda, k2k} and 
in solar electron neutrinos \cite{sksolar,sno}, as well as the 
KamLAND observation \cite{kamland} of 
electron antineutrino disappearance, are mainly due to neutrino 
oscillations 
and therefore neutrinos are massive.
One of the simplest ways to generate the small neutrino masses 
required is the seesaw mechanism \cite{seesaw}, in which three heavy singlet 
right-handed neutrinos, $\nu_R$, are added to the Standard Model (SM)
particle content. The Lagrangian giving rise to fermion masses is:
\beq
\mathcal{L} = \mathcal{L}_{SM} + Y_\nu \bar{l}_L H \nu_R + 
\frac 1 2 \bar{\nu}_R m_R \nu_R^c + {\rm h.c.},
\eeq
where $\mathcal{L}_{SM}$ stands for the usual SM Yukawa terms,
$l_L$ are the left-handed lepton doublets (we have suppressed family 
indices), $H$ is the Higgs doublet, $Y_\nu$ are the new Yukawa 
couplings that generate the neutrino Dirac masses after electroweak
symmetry breaking 
and $m_R$ is a 3 x 3 Majorana mass matrix.

   We expect the overall scale of $m_R$ to be much larger than the electroweak 
scale, because $\nu_R$ Majorana masses are 
$SU(3) x SU(2) x U(1)$ invariant and therefore naturally of the order of the 
cutoff of the low energy theory. Thus, we can integrate the $\nu_R$ away
and we get an effective Lagrangian which contains a Majorana mass term  
for the left-handed neutrinos,    
\beq
\mathcal{L}_{eff} = \frac 1 2 \bar{\nu}_L m_\nu \nu_L^c + {\rm h.c.}, 
\eeq
with
\beq
m_\nu \simeq m_D~m_R^{-1}~m_D^T
\label{seesaw}
\eeq
where $m_D=Y_\nu~v$ and $v$ is the VEV of the Higgs doublet.
If there is no quasi-degeneracy in the light neutrino masses, we can 
estimate the heaviest neutrino mass as 
\beq 
m_h \sim  \sqrt{\Delta m^2_{atm}} \simeq 5\times 10^{-2}\ eV \ ,
\eeq 
so for $v \lsim O(100 ~GeV)$ and $Y_\nu \sim O(1)$ 
the $\nu_R$ mass scale is $\Lambda \lsim  10^{15}~ GeV$, 
remarkably close to $M_{GUT}$.

  As a consequence of this high scale, lepton number violating 
scattering processes mediated by the heavy right-handed neutrinos
\beq
l_L+H \to \nu_R \to \bar{l}_L+\bar{H}
\label{scat}
\eeq
are in thermal equilibrium above the temperature $T \sim  10^{14}~ GeV$,
when the sphaleron interactions are not yet effective. 
In this new GUT baryogenesis scenario \cite{resurrection} the delayed
decay of coloured Higgs bosons produces baryon and lepton asymmetries,
conserving $B-L$, but the Majorana interactions erase the lepton 
number excess before the sphaleron interactions come into thermal
equilibrium, at $T_{sph} \sim 10^{12}~ GeV$. 
At lower temperatures, although the SM $B+L$ violating processes convert  
baryon number into lepton number, part of the original baryon
asymmetry survives, because now $B-L \neq 0$.

  In general, for this GUT baryogenesis mechanism to work we should
require that $\Gamma_{\Delta L = 2}(T_{sph}) < H(T_{sph})$, where 
\beq
 \Gamma_{\Delta L = 2}(T) = \frac{T^3}{4 \pi^3 v^4}\sum_{i=1}^{3}
m_{i}^2
\eeq
is the rate of the scattering processes
(\ref{scat})  and $H(T) \sim 17 T^2/M_{Plank}$ is the Hubble 
parameter at the temperature $T$.
Thus, if the light left-handed neutrinos are degenerate,   
since the rate of the lepton number violating scattering (\ref{scat})
is proportional to the sum of the squared neutrino masses,
the condition that it is out of equilibrium at $T \lsim 10^{12}~ GeV$
directly translates into an upper bound on the neutrino masses,
namely $m_{i} \lsim 0.2~eV$ for $v \sim 174~GeV$.

  Even if the lepton number violating scattering is out of equilibrium,
for right-handed neutrinos lighter than  $\sim 10^{12}~ GeV$
we should also require that their lepton number violating decays 
\beq
\nu_R \to \bar{l}_L+\bar{H}
\label{decay}
\eeq
are out of equilibrium at $T \lsim 10^{12}~ GeV$, otherwise 
these interactions together with the sphalerons would wash out 
the baryon asymmetry produced in the Higgs decays.
For each heavy neutrino, this decay rate is proportional to the mass parameter 
\beq
{\tilde m}_i = \frac{(Y^\dagger_\nu Y_\nu)_{ii}}{m_{\nu_{R_i}}}~ v^2 
\ ,\qquad  i=1,2,3
\eeq
where $m_{\nu_{R_i}}$ are the eigenvalues of the Majorana mass matrix $m_R$.
Notice that while the $\Delta L = 2$ scattering rate depends only on the 
light neutrino masses, the decay rate is model dependent.

Two comments are in order.
First, the constraint that the coloured Higgs should decay before the 
sphaleron processes enter into thermal equilibrium implies that 
$m_{H_c} > 10^{12}~ GeV$, which is in agreement with the limits 
from proton decay. 
Second, in principle this GUT  
baryogenesis mechanism  is also valid for 
supersymmetric models, however the high temperature at 
which the baryon asymmetry is produced ($T > 10^{12}~ GeV$)
may be a problem in inflationary cosmology, since it is well above 
the upper bound on the reheat temperature of the Universe from gravitino 
production. 

  We have addressed the question of whether this scenario for baryogenesis 
can be realised in several Grand Unified models of fermion masses
proposed in the literature. More precisely, we focus on models with 
horizontal symmetries in which neutrino masses are generated 
via the seesaw mechanism.
Our estimates are also applicable to the supersymmetric versions
of these models, with the caveat of the gravitino problem. 
For a more complete analysis and details, see \cite{us}.  
%

  \section{Models of Fermion Masses}

Atmospheric and solar neutrino oscillations can be fitted with two small
neutrino mass differences,
$\Delta m^2_{atm} = 2.5 \times 10^{-3}~eV^2$  and 
$\Delta m^2_{sol} = 6.9 \times 10^{-5}~eV^2$.
It is also known from experiment that the atmospheric mixing angle 
is nearly maximal, $c_{23} \sim s_{23} \sim 1/\sqrt{2}$, 
$s_{13} < 0.2$ according to CHOOZ \cite{chooz} and the solar 
angle is large.

These observations are consistent with three patterns 
of neutrino masses:
hierarchical ($\Delta m^2_{atm} \sim m_3^2$, 
$\Delta m^2_{sol} \sim m_2^2$), 
inverted hierarchical or quasi-Dirac 
($\Delta m^2_{atm} \sim m_3^2 \sim m_2^2$, 
$\Delta m^2_{sol} \sim m_3^2 - m_2^2$)
and degenerate ($m_3^2 \sim m_2^2 \sim m_1^2 \gg \Delta m^2_{atm}$).
The absolute scale is unknown, though, since neutrino oscillations are 
only sensitive to mass differences. 
In this respect, the search for neutrinoless double beta decay is 
crucial because it would provide information about the 
absolute spectrum and, moreover, discriminate between Majorana and 
Dirac neutrinos.

We assume three light Majorana neutrinos, whose masses arise from the 
minimal seesaw mechanism. As we have seen, in this context the 
scale $\Lambda$ for $L$ non-conservation is suggestively close to 
$M_{GUT}$, so we consider Grand Unified models for all 
fermion masses.   
Horizontal (or family) symmetries have been proposed with the aim of 
explaining the hierarchical pattern of quark and charged lepton masses
\cite{U1}.
 For simplicity, we have analysed models with abelian $U(1)_H$ 
family symmetry.

The horizontal symmetry allows only third family (or just top quark) Yukawa 
couplings, however it is spontaneously broken 
by the vev's $v_f$ of some ``flavon'' fields $\Phi$, which are singlets 
under the grand unified gauge group but charged under the flavour symmetry.
Then smaller couplings 
are generated effectively from higher dimension non-renormalisable operators 
but are suppressed by powers of the parameter $v_f / \Lambda_c$, where 
 $\Lambda_c$ is the cut-off of the theory. We expect 
$v_f \gtrsim M_{GUT}$ and $\Lambda_c \lsim M_{Plank}$, so the 
ratio $v_f / \Lambda_c$ can be identified with the 
Wolfenstein parameter, $\lambda \sim 0.22$.
The number of powers of the expansion parameter is controlled by the 
flavor charge of the particular operator.

Notice that in all the models considered (and in most of this kind existent 
in the literature) the light neutrino masses are hierarchical, since it
is difficult to obtain a degenerate spectrum in the context of GUT 
models of fermion masses.
Then, as we explained before, $\Delta L = 2$ scattering processes are 
always out of equilibrium at $T_{sph} \sim 10^{12}~ GeV$, when 
the $B-L$ conserving interactions become active.
So what we require for the modified GUT baryogenesis to work is that 
either the right-handed neutrinos are heavier than $10^{12}~ GeV$,
and thus they are not present in the thermal bath when the sphalerons 
enter into thermal equilibrium or, if they are lighter, their 
 $\Delta L = 1$ decays are out equilibrium at $T \leq T_{sph}$,
i.e., 
$\Gamma_{\Delta L = 1}(T=M_i) < H(T=M_i)$ for $M_i \leq 10^{12}~ GeV$
\footnote{Note that this condition ensures that 
$\Gamma_{\Delta L = 1}(T) < H(T)$ at any $T>M_i$, so it is enough to
consider $T=M_i$.}.

\subsection{$SU(5)\times U(1)_H$ models}
\label{su5}

In Grand Unified $SU(5)$ models the SM quarks and leptons
 are grouped in the multiplets ${\bf 10}=(q_L,u_R^c,e_R^c)$ and 
${\bf 5^*}=(d_R^c,l_L)$ and the right-handed neutrinos are added in the singlet
 representation ${\bf 1}=\nu_R^c$. 
The Higgs multiplets transform as a ${\bf 5}$, and 
contain the coloured triplets whose 
decays generate the baryon asymmetry in the proposed mechanism.
We assume that there are two Higgs multiplets because it has been shown that 
with just one the baryon asymmetry generated is too small 
\cite{nanwei-barr}.

	\subsubsection{Q(fermions) $\geq 0$}

       First we consider a model in which the $U(1)_H$ symmetry is broken 
by one singlet field $\Phi$ with charge $Q = - 1$, and all fermions with 
non-negative charges \cite{buchyan}. The expansion parameter 
in this model is chosen to be  
$\epsilon = \langle \Phi \rangle / \Lambda_c \simeq 0.06$ i.e., 
$\epsilon \simeq \lambda^2$. 
 From the observed hierarchy of quark and charged lepton masses one can 
deduce that the 
allowed $U(1)_H$ charges are 
\beq
\renewcommand{\arraystretch}{1.5}
\begin{array}{|ccc|ccc|ccc|} 
\hline
{\bf 10_3}&{\bf 10_2}&{\bf 10_1}&
{\bf 5_3^*}&{\bf 5_2^*}&{\bf 5_1^*}&{\bf 1_3}&{\bf 1_2}&{\bf 1_1}\cr
\hline
\hline
0&1&2&a&a&a+1&b&c&d\cr
\hline 
\end{array}
\eeq
where the value of the b quark mass allows $a=0,1$ \cite{vala} 
and $0\le b\le c \le d$.

	With those charges one can readily calculate the Dirac neutrino
mass matrix and the Majorana mass matrix of the right-handed neutrinos,
up to order one coefficients \cite{buchyan}.
Then,   
using the seesaw formula (\ref{seesaw}) one obtains the effective Majorana 
mass matrix for the light neutrinos:
\beq
m_\nu=\epsilon^{2a}~\matriz{
\epsilon^2&\epsilon&\epsilon\\
\epsilon&1&1\\
\epsilon&1&1}~\frac{v^2}{\Lambda}
\label{mnu1}
\eeq

Notice that in this kind of models with one single flavon field and all
fermions with non-negative charges the dependence on the $U(1)_H$ charges of
the right-handed neutrinos drops out, and the light neutrino 
Majorana mass matrix depends only on the charges of the 
{\bf 5} fermions.
Thus it has always the hierarchy structure of (\ref{mnu1}), that
has been proposed (among others that we will discussed later) 
to generate naturally the large atmospheric mixing angle.
The hierarchy between $m_2$ and $m_3$ is then considered accidental,
since the sub-determinant 23 of $m_\nu$ is in general $O(1)$,
but this is not unlikely for the LMA solution of the solar neutrino 
problem: $m_2/m_3 \sim 0.1$ is not particularly small and could 
easily arise from a combination of order one terms.

We have analysed different values of the $U(1)_H$ charges $a,b,c,d$.
Clearly, the modified GUT baryogensis scenario requires heavy right-handed 
neutrinos so the most favourable case is $a=b=c=d=0$, 
which corresponds to almost degenerate right-handed neutrinos 
 with masses of order $\Lambda \sim 10^{14}~ GeV$. Therefore 
for these $U(1)_H$ charges the mechanism is feasible. 

The cases $a=1, b=c=d=0$
and $a=b=c=0, d=1$	
are somehow in the limit because the lightest right-handed neutrino masses 
are $M_i \sim 10^{12}~ GeV$ and we have found that 
$\Gamma_{\Delta L = 1}(T=M_i) \gsim H(T=M_i)$. Thus a more detailed 
analysis would be needed to elucidate these cases.

For all other values of the $U(1)_H$ charges, the 
lightest right-handed neutrino mass is well bellow $10^{12}~ GeV$
and its $\Delta L = 1$ decays are in equilibrium at $T \lsim T_{sph}$
so GUT baryogenesis is not possible.

         \subsubsection{Q(fermions) of both signs}
 
 One can also consider that the $U(1)_H$ family symmetry is broken by a 
pair of $SU(5)$ singlets $\Phi$ and $\bar{\Phi}$ with charges $Q=1$ and
$Q=-1$, respectively \cite{AF}. Taking 
$\lambda = \langle \Phi \rangle / \Lambda_c \sim
\lambda' = \langle \bar{\Phi} \rangle / \Lambda_c \sim 0.22$, 
the fermion masses can be described by the following set of charges:
\beq
\renewcommand{\arraystretch}{1.5}
\begin{array}{|ccc|ccc|ccc|} 
\hline
{\bf 10_3}&{\bf 10_2}&{\bf 10_1}&
{\bf 5_3^*}&{\bf 5_2^*}&{\bf 5_1^*}&{\bf 1_3}&{\bf 1_2}&{\bf 1_1}\cr
\hline
\hline
0&2&3&0&0&b&0&-a&a\cr
\hline 
\end{array}
\eeq
where  $b \ge 2 a >0$. If $b$ = 2 or 3, the up, down and charged lepton sectors
are not essentially different from the previous case. In the neutrino 
sector, the Dirac and Majorana mass matrices are given by
\beq
m_D=\matriz{
\lambda^{a+b}&\lambda^{b-a}&\lambda^{b}\\
\lambda^{a}&  \lambda'^{a}& 1 \\
\lambda^{a}&  \lambda'^{a}&  1}~v
\qquad \qquad
m_R=\matriz{
\lambda^{2a}&1&\lambda^{a}\\
1&\lambda'^{2a}&\lambda'^{a}\\
\lambda^{a}&\lambda'^{a}&1}~\Lambda
\eeq
The $O(1)$ off-diagonal entry of $m_D$ is typical of the so-called
lopsided models, which have been proposed to obtain simultaneously 
the large atmospheric neutrino mixing and the observed mass splitting
between the solar and atmospheric frequencies.
After diagonalising the charged lepton sector and integrating out the 
heavy right-handed neutrinos, one obtains the effective Majorana 
mass matrix for the light neutrinos:
\beq
m_\nu=\matriz{
\lambda^{2b}&\lambda^b&\lambda^b\\
\lambda^b& 1+ \lambda^a \lambda'^a& 1 + \lambda^a \lambda'^a\\
\lambda^b &1+\lambda^a \lambda'^a&1+\lambda^a \lambda'^a}
~\frac{v^2}{\Lambda}
\label{mnu2}
\eeq

The important property of $m_\nu$ is that, as a result of the seesaw 
mechanism and the $U(1)_H$ charge assignments, the determinant of the 
23 block is automatically of $O(\lambda^a \lambda'^a)$, while 
the $O(1)$ elements lead to large atmospheric mixing angle.
This property is entirely due to the specific lopsided form of the Dirac 
neutrino mass matrix $m_D$, for any generic Majorana matrix 
$m_R$, with the only condition that the 33 entry is non vanishing, 
and that no new $O(1)$ terms are generated in $m_\nu$ by a 
compensation between small terms in $m_D$ and large terms in 
$m_R^{-1}$. 

For $\lambda \sim  \lambda'$, it is easy to verify that the 
eigenvalues of $m_\nu$ satisfy the relation
$m_1 : m_2 : m_3 = \lambda^{2(b-a)}:\lambda^{2a}:1$
and that the solar mixing angle is 
$\theta_{12} \sim \lambda^{b-2a}$. Thus, it is possible to reproduce the MSW 
large mixing angle solution for $b=2$ and $a=1$.

Again, in this type of models the three right-handed neutrinos are 
heavy, with masses of order $\Lambda \sim 10^{14}~ GeV$, and  
the GUT baryogenesis seems viable.

	\subsection{$SO(10)$ models}

   It is well known that the three SM families of quarks and leptons
together with the extra right-handed neutrinos fit neatly
into three copies of the $SO(10)$ spinor representation {\bf 16}. 
This feature has made $SO(10)$ one of the most attractive 
unification groups.
  Regarding $SO(10)$ Grand Unified models we first consider a detailed 
top-down model, that fits all fermion masses and mixings, and then we follow a 
bottom-up approach, which aims to reconstruct the 
right-handed neutrino Majorana matrix $m_R$ from the low energy
effective mass matrix $m_\nu$, in the framework of a $SO(10)$ inspired pattern
for the Dirac neutrino mass matrix, $m_D$. 
In both cases, neutrino masses are generated via the minimal seesaw,
without additional Majorana mass terms due to small vevs of $SU(2)$ triplets,
which are possible in $SO(10)$ models \cite{triplets}.

	\subsubsection{Albright-Barr model}
	
	This model, developed in 
\cite{albright-barr}, is based on a $SO(10)$ gauge group supplemented by a 
horizontal symmetry $U(1)_H \times Z_2 \times Z_2$.
It is an attempt to construct a realistic $SO(10)$ model with the minimal 
Higgs content.
The Dirac mass matrices for the neutrinos and 
charged leptons are 
\beq \label{matrSO10}
m_D=Y_\nu~M_U = \matriz{\eta&0&0\\0&0&-\epsilon\\0&\epsilon&1}~M_U
\qquad \qquad 
m_l=Y_l~M_D=\matriz{\eta&\delta&\delta'e^{i\phi}\\
\delta&0&-\epsilon\\\delta'e^{i\phi}&\sigma+\epsilon&1}~M_D 
\eeq
The zeroes arise from restrictions because of the 
$U(1) \times Z_2 \times Z_2$ flavour symmetry, which forbids 
any Froggatt-Nielsen diagram. 
The antisymmetric $\epsilon$ terms arise from diagrams involving the 
adjoint $\langle \bf{45}_H \rangle$ Higgs vev pointing in the 
$B-L$ direction. 
The Dirac mass matrices for the quarks can be found in  
\cite{albright-barr}.

The model fits all nine quark and charged lepton masses plus the three CKM 
angles and CP phase with the eight input parameters 
\beq
\begin{array}{ccc}
M_U \simeq 113~GeV &&M_D \simeq 1~GeV\\
\sigma = 1.78 && \epsilon = 0.145\\
\delta = 0.0086 && \delta' = 0.0079\\
\phi = 54^o && \eta = 8 \times 10^{-6}
\end{array}
\label{inputs}
\eeq

As a consequence of the lopsided nature of the large 
$\sigma$ term in the matrix $m_l$, the hermitian matrix $m_l^\dagger m_l$
is diagonalised by a large left-handed rotation, which accounts for 
the near maximal atmospheric neutrino mixing for any reasonable 
neutrino Majorana mass matrix $m_R$.

	The type of solar neutrino mixing is determined 
by the texture of $m_R$, which is rather independent of the Dirac mass 
matrices (\ref{matrSO10}) because it arises from 
completely different operators \cite{albright-barr}. 
In particular, the LMA solution of the solar neutrino 
problem requires a nearly hierarchical texture, namely 
\beq
m_R=\matriz{c^2\eta^2&-b\epsilon\eta&a\eta\\
-b\epsilon\eta&d^2\epsilon^2&-d \epsilon\\
a\eta&-d \epsilon&1\\}~\Lambda
\eeq
where $\epsilon$ and $\eta$ are specified in eq. (\ref{inputs})
and $a, b, c, d$ are order 1 parameters restricted by neutrino data.
The fact that the same $d$ appears in the 22, 23 and 32 elements 
of $m_R$ is due to the factorised structure of the diagrams leading to
the right-handed neutrino masses, and 
therefore is not fine-tuning. However this coefficient is 
strongly constrained by the CHOOZ bound, 
$s_{13} \lsim 0.2$, which requires 
$0.85 \leq d \leq 1.15$ \cite{albright-barr}.

Using the seesaw formula with $d=1$ one obtains the 
effective left-handed neutrino Majorana mass matrix 
\beq
m_\nu=\matriz{0&\epsilon/(a-b)&0\\
\epsilon/(a-b)&-\epsilon^2(c^2-b^2)/(a-b)^2&-b \epsilon/(a-b)\\
0&-b \epsilon/(a-b)&1\\}~M_U^2/\Lambda \ ,
\label{mnu4}
\eeq
which leads to a normal hierarchical spectrum for the light neutrinos.

One of the simplest cases allowed is $a=1$, $b=c=2$ and 
$\Lambda=10^{14}~GeV$. 
Then, the two lightest right-handed neutrinos have masses of order 
$m_{\nu_{R_1}}\sim m_{\nu_{R_2}} \sim 10^8~GeV$ and we have found that 
their $\Delta L \neq 0$ decays are in equilibrium below $10^{11}~GeV$,
so together with the sphaleron processes they will wash out 
the baryon asymmetry generated during the Higgs decays.
Given the strong hierarchy of $m_R$, we expect that this result is 
generic for any value of the free parameters of the model.

	\subsubsection{Branco \etal~ approach}

	The bottom-up approach used in \cite{branco} 
predicts the $\nu_R$ masses from $\nu_L$ parameters.
The $m_\nu$ matrix can be written as 
$m_\nu = U_{MNS}~d_\nu~U_{MNS}^T$, 
where we know the mass squared differences and the mixing angles.
In the basis where the charged lepton mass matrix $m_l$ is real and 
diagonal, the Dirac neutrino mass matrix can be written as 
$m_D=V_L^\dag~d_D~U_R$, with $d_D$ diagonal and $V_L, U_R$  
unitary matrices.
Then, using the seesaw formula
\beq
m_\nu=m_D~m_R^{-1}~m_D^T=-V_L^\dag~d_D~U_R~m_R^{-1}~U_R^T~d_D~V_L^*
\eeq
The $V_L$ matrix is taken to be the identity 
in the analytic approach of \cite{branco}; this is a reasonable 
approximation in a minimal $SO(10)$ scenario, where one expects $V_L$ to be 
of the order of the CKM matrix, although it may not be the case  
in more realistic models.   
Defining
 $M\equiv U_R~m_R^{-1}~U_R^T$ so
\beq
m_\nu \simeq -d_D~M~d_D
\eeq
one gets 
\beq
M \simeq -d_D^{-1}~m_\nu~d_D^{-1}
\eeq
	Note that $M$ is  $m_R^{-1}$ 
in the basis where the Dirac mass matrix $m_D$ is diagonal.
Now, using the  $SO(10)$ motivated relation $m_D \sim m_u$
and taking into account the up-quark mass hierarchy at the 
GUT scale, one has
\beq
d_D \sim \matriz{m_u&0&0\\0&m_c&0\\0&0&m_t} \simeq 
\matriz{\epsilon^2&0&0\\0&\epsilon&0\\0&0&1\\}~m_t
\eeq
where $\epsilon \sim \lambda^4 \simeq 3 \times 10^{-3}$. 

The matrix $M$ can be analytically diagonalised at leading order in 
$\epsilon$ (see ref. \cite{branco} for details), and the right-handed 
neutrino masses are just the inverse of these eigenvalues.
As in the previous $SO(10)$ model, the $\nu_R$ have a strong hierarchy:
$m_{\nu_{R_1}}:m_{\nu_{R_2}}:m_{\nu_{R_3}} \sim \epsilon^4:\epsilon^2: 1$, 
so we expect at least one of the right-handed neutrinos to be lighter 
than $T_{sph} \sim 10^{12}~ GeV$ and, as a consequence, it is unlikely 
that the GUT baryogenesis works within this framework.

For instance, assuming a hierarchical spectrum of the light left-handed 
neutrinos, $m_{1} \ll m_{2} \ll m_{3}$, 
we found
\beq
m_{\nu_{R_1}} \sim \frac{m_t^2~\epsilon^4}{m_2~\sin^2 \theta_{12}} 
\sim 10^5~GeV
\qquad
m_{\nu_{R_2}} \sim \frac{2~m_t^2~\epsilon^2}{m_3} \sim 10^9~GeV\qquad
m_{\nu_{R_3}} \sim \frac{m_t^2~\sin^2 \theta_{12}}{2~m_1}
\eeq
where we have neglected $s_{13}$ and we have taken $\theta_{12}\sim \pi/4$, 
$m_3 = \sqrt{\Delta m^2_{atm}}\simeq 5\times 10^{-2}~eV$ and 
$m_2 =\sqrt{\Delta m^2_{sol}} \simeq 8\times 10^{-3}~eV$.

	We have calculated that the $\Delta L =1$
decays of the lightest right-handed neutrinos
$\nu_{R_{1,2}}$ are in  equilibrium around  $10^6~GeV$ and $10^9~GeV$ 
respectively,  
so those decays continue destroying lepton number at $T\lsim 10^{12}~GeV$ 
when the sphaleron configurations are also in equilibrium.
Then, the baryon asymmetry generated in the coloured Higgs decays 
is completely washed out.
Similar results are obtained if we assume an inverted-hierarchical 
spectrum of the left-handed neutrinos.

\subsection{Pati-Salam}
\label{ps}

In this section we analyse a supersymmetric string-inspired model of 
all fermion masses and mixing angles based on the Pati-Salam 
$SU(4)\times SU(2)_L \times SU(2)_R$ gauge group \cite{patisalam} supplemented 
by a $U(1)_H$ flavour symmetry \cite{kingps}.
It provides an example of a realistic grand unified model in which both the 
large atmospheric mixing angle and the hierarchical neutrino masses arise
naturally, due to   
the so-called single right-handed neutrino dominance (SRHND) mechanism,
that we will discuss in the next section.
Here we are just concerned about the neutrino sector, see ref.\cite{kingps}
for details of the full model.

  An attractive feature of grand unified models based on the Pati-Salam
group is that the presence of the gauged $SU(2)_R$ subgroup 
predicts the existence of the right-handed neutrinos, much as in $SO(10)$. 
 The SM fermions, together with the right-handed neutrinos, are
accommodated  in the $F=(4,2,1)$ and the $F^c=(\bar{4},1,\bar{2})$
representations, 
\beq
F_A = \left( \begin{array}{cccc}
u&u&u&\nu\\d&d&d&e\end{array} \right)_A  \qquad \qquad
F_B^c = \left( \begin{array}{cccc}
d^c&d^c&d^c&e^c\\u^c&u^c&u^c&\nu^c \end{array} \right)_B  
\eeq
and the SM Higgs boson is contained in $h=(1,\bar{2},2)$.
  In order to obtain a realistic SM fermion spectrum, 
the mass matrices are generated by non-renormalisable operators of the 
form
\beq
F_A F_B^c h 
\left(\frac{H \bar{H}}{\Lambda_c^2}\right)^n 
\left(\frac{\theta}{\Lambda_c} \right)^{p_{AB}}
\qquad {\rm and} \qquad
F_A F_B^c h 
\left(\frac{H \bar{H}}{\Lambda_c^2}\right)^n 
\left(\frac{\bar{\theta}}{\Lambda_c} \right)^{p_{AB}}
\eeq
where $H=(4,1,2)$ and $\bar{H}=(\bar{4},1,\bar{2})$ 
are the heavy Higgs fields that break the 
$SU(4)\times SU(2)_L \times SU(2)_R$ symmetry 
to the SM group at $M_{GUT}$, 
$\theta, \bar{\theta}$
are the Pati-Salam singlets which carry $U(1)_H$ charges $\pm 1$ 
and break the flavour symmetry, and $\Lambda_c > M_{GUT}$ denotes the mass
of extra matter that has been integrated out.
When the $H$ and $\theta$ fields develop their vev's, those 
operators reduce to effective Yukawa couplings with powers of the 
small coefficients 
\beq
\delta = \frac{\langle H \rangle \langle \bar{H} \rangle}{\Lambda_c^2} 
\simeq 0.22 
\qquad \qquad \epsilon = \frac{\langle \theta \rangle }{\Lambda_c} \simeq 0.22
\eeq
Then one can identify $\delta$ with mass splitting within generations 
and $\epsilon$ with splitting between generations.
On the other hand, the right-handed neutrinos $\nu^c$ acquire a large mass, 
$(m_R)_{ij} \sim \delta^m \epsilon^{q_{ij}} \langle H H \rangle/\Lambda_c$.

	This model predicts the following mass matrices
\beq
m_D
\sim 
\matriz{\delta^3 \epsilon^5&\delta \epsilon^3&\delta \epsilon\\
\delta^3 \epsilon^4&\delta^2 \epsilon^2&\delta\\
\delta^3 \epsilon^4&\delta^2 \epsilon^2&1\\}~ v \sim 
\matriz{ {\lambda}^8&{\lambda}^4& {\lambda}^2\\ 
{\lambda}^7& {\lambda}^4& {\lambda}\\ {\lambda}^7& {\lambda}^4&1\\}~ v
\eeq\\
\beq
m_R
\sim 
\matriz{\delta \epsilon^8&\delta \epsilon^6&\delta \epsilon^4\\
\delta \epsilon^6&\delta \epsilon^4&\delta \epsilon^2\\
\delta \epsilon^4&\delta \epsilon^2&1\\}~ \Lambda \sim 
\matriz{ {\lambda}^9& {\lambda}^7& {\lambda}^5\\ 
{\lambda}^7& {\lambda}^5& {\lambda}^3\\ {\lambda}^5& {\lambda}^3&1\\}~ \Lambda
\eeq\\

	Then, using the seesaw mechanism (\ref{seesaw}) one can calculate 
the effective Majorana mass matrix $m_\nu$,
\beq
m_\nu \sim \matriz{ {\lambda}^2& {\lambda}^{3/2}& {\lambda}^{3/2}\\ 
{\lambda}^{3/2}&1&1\\ 
{\lambda}^{3/2}&1&1\\}~ \frac{v^2}{\Lambda} \ ,
\label{mnu6}
\eeq
which is dominated by the third right-handed neutrino.

Requiring that the heaviest mass eigenvalue is 
$m_3 = \sqrt{\Delta m^2_{atm}} \simeq 5\times 10^{-2}\ eV$, 
the mass scale of the right-handed neutrinos results to be 
$\Lambda \lsim 6 \times 10^{14}~GeV$  so the two lightest $\nu_R$'s have  
masses $m_{\nu_{R_1}} \lsim 7 \times 10^{8}~GeV$, 
$m_{\nu_{R_2}} \lsim 3 \times 10^{11}~GeV$.
Moreover, we have calculated that their $\Delta L =1$ decays are in 
equilibrium around $10^{9}~GeV$ for $\nu_{R_1}$ 
and $7 \times 10^{11}~GeV$ for $\nu_{R_2}$, so  these decays continue 
destroying lepton number at $T\lsim 10^{12}~GeV$, with the sphaleron 
configurations in equilibrium, and the baryon asymmetry 
that could be generated in the heavy Higgs decays disappears.
Therefore the GUT baryogenesis mechanism does not work within this model.

	\subsection{SRHND}

\label{srhnd}

	The single right-handed neutrino dominance (SRHND) \cite{king} 
is an alternative mechanism proposed in the literature to generate 
simultaneously a large atmospheric mixing angle, $\theta_{23}$, and 
hierarchical left-handed neutrino masses. 
The idea is that the dominant contributions to the 23 block of the 
light effective Majorana mass matrix $m_\nu$ come from only one of the 
right-handed neutrinos, leading to an approximately vanishing  
23 sub-determinant. 
Next to leading contributions from the other 
right-handed neutrinos are required to generate the mass splitting
and large 12 mixing angle corresponding to the LMA solution 
of the solar neutrino puzzle.

This mechanism can be realised in many different models, in particular 
it has been thoroughly studied in the framework of $U(1)_H$ 
flavor symmetry models. 
These can be embedded in grand unified theories by imposing the 
corresponding constraints on the quark and lepton $U(1)_H$ charges,
although usually the minimal models are difficult to reconcile with 
the data. Therefore in order to analyse if GUT baryogenesis is 
possible when SRHND takes place,
we shall just study different charge assignments 
to the leptonic sector which reproduce the measured neutrino mass splittings 
and mixing angles, without specifying the quark charges 
corresponding to a particular grand unified group.

  If we take $\nu_{R_3}$, of mass $Y$, to be the dominant right-handed 
neutrino, there are three possible textures of the heavy Majorana mass matrix 
$m_R$ that maintain its isolation from the other right-handed  
neutrinos, namely the diagonal, democratic and off-diagonal 
textures \cite{king} 
\beq \label{mR}
\begin{array}{c}
m_R^{diag}=\matriz{
X'&0&0\\
0&X&0\\
0&0&Y}
\qquad
m_R^{dem}=\matriz{
X&X&0\\
X&X&0\\
0&0&Y}
\qquad
m_R^{off-diag}=\matriz{
0&X&0\\
X&0&0\\
0&0&Y}
\end{array}
\eeq
In the democratic and off-diagonal textures, the Majorana masses of the upper 
block, $X$, are of the same order but not exactly equal.

	Denoting the Dirac neutrino mass matrix $m_D$ as
\beq \label{mD}
m_D=\matriz{
a'&a&d\\
b'&b&e\\
c'&c&f}\ v
\eeq
 and using the seesaw formula, one obtains the effective light neutrino 
Majorana mass matrix, $m_\nu$.
The conditions for SRHND are that the $1/Y$ terms in the 23 sub-matrix 
dominate over the $1/X, 1/X'$ terms in the full matrix,
\beq
\frac{e^2}{Y} \sim \frac{e f}{Y} \sim \frac{f^2}{Y} \gg
\mathcal{O}\left(\frac{1}{X}\right), \mathcal{O}\left(\frac{1}{X'}\right)
\eeq

At order $\mathcal{O}\left(\frac{1}{Y}\right)$
the matrix $m_\nu$ is the same for the three textures, leading to
the heaviest neutrino mass 
\beq
m_{\nu_3}=\frac{d^2+e^2+f^2}{Y}v^2
\eeq
and the mixing angles  \cite{king}
\beq
\tan \theta_{23}=\frac{e}{f} 
\qquad \qquad 
\tan \theta_{13}=\frac{d}{\sqrt{e^2+f^2}}
\eeq
Taking into account that $\tan \theta_{23} \approx 1$ from Super-Kamiokande
and $\tan \theta_{13} \lsim 0.2$ from CHOOZ 
\cite{chooz}, it is necessary that $d\ll e\approx f$  \cite{king}.

 The masses of the two remaining neutrinos as well as the solar mixing 
angle $\theta_{12}$ appear at subleading order, and their 
analytic expressions depend on the particular texture of $m_R$.
In the diagonal case, if $|x y|/X \gg |x' y'|/X'$ 
($x,y =a,b,c$ and $x',y' =a',b',c'$),
 we have $m_{\nu_1} \ll m_{\nu_2}$, while
 the diagonal texture when both $|x y|/X$ and $|x' y'|/X'$ terms are 
important, as well as the other two textures, lead to 
$m_{\nu_1} \sim  m_{\nu_2}$.

 In \cite{king}
 it has been shown that the conditions for SRHND and LMA solution 
of the solar neutrino deficit can be fulfilled in
the framework of models with $U(1)_H$ family symmetries, 
with appropriate charge assignments, and a 
systematic search of the simplest examples has been performed.
They assume that the $U(1)_H$ is broken by the equal vevs of two 
singlets, $\Phi,~\bar{\Phi}$, which have charges $\pm 1$, 
so both signs for the charges of the leptons are allowed. 
Then, omitting order one coefficients as usual, the Dirac neutrino 
mass matrix and the heavy Majorana matrix are
\beq
m_D=\matriz{
{\lambda}^{|n_1+l_1|}& {\lambda}^{|n_2+l_1|}& {\lambda}^{|n_3+l_1|}\\
 {\lambda}^{|n_1+l_2|}& {\lambda}^{|n_2+l_2|}& {\lambda}^{|n_3+l_2|}\\
 {\lambda}^{|n_1+l_3|}& {\lambda}^{|n_2+l_3|}& {\lambda}^{|n_3+l_3|}\\}~v
\eeq
\beq
m_R=\matriz{
 {\lambda}^{|2 n_1+\sigma|}& {\lambda}^{|n_2+n_1+\sigma|}& 
{\lambda}^{|n_3+n_1+\sigma|}\\
 {\lambda}^{|n_1+n_2+\sigma|}& {\lambda}^{|2 n_2+\sigma|}& 
{\lambda}^{|n_3+n_2+\sigma|}\\
 {\lambda}^{|n_1+n_3+\sigma|}& {\lambda}^{|n_2+n_3+\sigma|}& 
{\lambda}^{|2 n_3+\sigma|}\\} \Lambda
\eeq
where the charges $l_i,~n_i,~\sigma$ correspond to the $l_L,~\nu_R$ and $S$ 
fields,
being $S$ the one whose vev generates the heavy Majorana neutrino masses, 
and $\lambda\sim 0.22$ is the Wolfenstein parameter, given by the 
ratio $\langle \Phi \rangle /\Lambda_c = \langle \bar{\Phi}\rangle /\Lambda_c$.
It is important for SRHND that at least some of the combinations  
$l_i + n_p$ and $n_p + n_q + \sigma$ take negative values; otherwise, 
the modulus signs could be dropped and the right-handed neutrino charges
$n_p$ would cancel in $m_\nu$, as constructed using the seesaw 
formula (\ref{seesaw}). Therefore the choice of right-handed neutrino charges 
plays an important role in determining $m_\nu$ and each particular 
choice must be analysed separately.

  We have studied several of the $U(1)_H$ charge assignments found in 
\cite{king}, which provide a natural account of the atmospheric and 
solar neutrinos via the LMA MSW effect. We have focused on choices 
that lead to all right-handed neutrino masses $m_{\nu_{R_i}}>10^{12}~GeV$,
so they are not present in the thermal bath when the sphaleron 
processes enter into thermal equilibrium and GUT baryogenesis is 
in principle feasible
\footnote{The possibility of lighter right-handed neutrinos but with 
$\Delta L =1$ decays out of equilibrium at $T \lsim 10^{12}~ GeV$
will be considered in \cite{us}.}. 
For example, for diagonal textures we found that in the following cases
GUT baryogenesis is viable
\begin{equation}
\renewcommand{\arraystretch}{1.3}
\begin{array}{|ccc|ccc|c|}\hline
l_1 & l_2 & l_3 & n_1 & n_2 & n_3 & \sigma \\ \hline\hline
-1 & 1 & 1 & 0 & \frac{1}{2} & -\frac{1}{2} & -1 \\ \hline
-1 & 1 & 1 & \frac{1}{2} & 0 & -\frac{1}{2} & -1 \\ \hline
-\frac{1}{2} & \frac{1}{2} & \frac{1}{2} & \frac{1}{2} & 0 & -1 & -1 \\ \hline
\end{array} \ ,
\end{equation} 
and as working examples of the democratic and the off-diagonal textures we have
\footnote{Note that the third example of off-diagonal textures can be readily 
embedded in a grand unified $SU(5)$ theory, like the second one we considered 
in section \ref{su5}.}
\begin{equation}
\renewcommand{\arraystretch}{1.3}
\begin{array}{|ccc|ccc|c|}\hline
l_1 & l_2 & l_3 & n_1 & n_2 & n_3 & \sigma \\ \hline\hline
-1 & 0 & 0 & \frac{1}{2} & \frac{1}{2} & -1 & -1 \\ \hline
-1 & 0 & 0 & \frac{1}{2} & \frac{1}{2} & -\frac{1}{2} & -1 \\ \hline
-1 & 0 & 0 & \frac{1}{2} & \frac{1}{2} & 0 & -1 \\ \hline
\end{array} 
\qquad \qquad
\begin{array}{|ccc|ccc|c|}\hline
l_1 & l_2 & l_3 & n_1 & n_2 & n_3 & \sigma \\ \hline\hline
-2 & 0 & 0 & -2 & 1 & -1 & 1 \\ \hline
-2 & 0 & 0 & -2 & -2 & 0 & 0 \\ \hline
-2 & 0 & 0 & -1 & 1 & 0 & 0 \\ \hline
\end{array} 
\end{equation}

	\subsection{Inverted hierarchy models}

	This kind of models have been studied in \cite{inverted} and 
are based on a mechanism similar to SRHND, but with dominance of two heavy 
right-handed neutrinos. A reversal of the SRHND conditions leads to an 
inverted hierarchical spectrum of the left-handed neutrinos,  
$m_{\nu_3} \ll m_{\nu_2}\sim m_{\nu_1}$, and bi-maximal mixing. 
As in the case of SRHND, the mechanism can naturally take place in 
models with a $U(1)_H$ family symmetry broken by two vector-like singlets 
with charges $\pm 1$. 
We shall only be concerned about the leptonic sector, although 
these models could be extended to describe the quark masses
as well, in a GUT scenario.

The only texture of the heavy Majorana neutrino matrix, $m_R$, that 
can lead to an inverted mass hierarchy  
is the off-diagonal one in eq. (\ref{mR}) \cite{inverted}.
  Using for the Dirac neutrino mass matrix $m_D$ the same notation as in 
section \ref{srhnd} (see eq. (\ref{mD})), the seesaw formula (\ref{seesaw})
implies
\beq
m_\nu=\matriz{
\frac{d^2}{Y} + \frac{2 a a'}{X}&\frac{d e}{Y}+\frac{a'b+ab'}{X}&
\frac{d f}{Y}+\frac{a'c+ac'}{X}\\
\frac{d e}{Y}+\frac{a'b+ab'}{X}&\frac{e^2}{Y}+\frac{2bb'}{X}&
\frac{e f}{Y}+\frac{b'c+cb'}{X}\\
\frac{d f}{Y}+\frac{a'c+ac'}{X}&\frac{e f}{Y}+\frac{b'c+cb'}{X}&
\frac{f^2}{Y}+\frac{2cc'}{X}\\}\ v^2
\label{mnu8}
\eeq
In order to achieve an inverted hierarchy one should 
require that 
the $\mathcal{O}\left(\frac{1}{X}\right)$ terms dominate 
over the $\mathcal{O}\left(\frac{1}{Y}\right)$ ones 
(which is the opposite of the SRHND condition), and that either 
$a',b,c \gg a,b',c$ or $a',b,c \ll a,b',c'$. 
In the first case one finds the following $\nu_L$ masses
\beq
\begin{array}{c}
{\displaystyle
m_{\nu_3} \approx\frac{(b-c)^2}{b^2+c^2}\ \frac{1}{Y}\ v^2 \qquad \qquad 
m_{\nu_1} \approx m_{\nu_2} \approx \frac{a'\sqrt{b^2+c^2}}{X}\ v^2 }
\end{array} \ .
\eeq

	Introducing a $U(1)_H$ flavour symmetry, the conditions for 
achieving the required off-diagonal texture of $m_R$ and an inverted 
hierarchical spectrum of the left-handed neutrinos
translate into conditions of the $U(1)_H$ charges for the different fields.
Many working examples can be found in \cite{inverted}, of which we have 
analysed some. In particular, we have obtained that the three right-handed 
neutrinos are heavier than $10^{12}~ GeV$ and therefore GUT baryogensis 
is viable for the following choices 
of the $U(1)_H$ charges 
\begin{equation}
\begin{array}{|ccc|ccc|c|}\hline
l_1 & l_2 & l_3 & n_1 & n_2 & n_3 & \sigma \\ \hline\hline
-2 & 3 & 3 & 2 & -2 & 0 & 0 \\ \hline
-3 & 1 & 1 & 3 & -2 & 1 & -2 \\ \hline
-3 & 3 & 3 & 3 &-3 & -1 & 1 \\ \hline
\end{array} 
\end{equation} 
where the charges $l_i,~n_i,~\sigma$ correspond to the $l_L,~\nu_R,~S$ fields,
as in the case of SRHND.


  \section{Conclusions}

  We have found that the GUT-baryogenesis mechanism 
proposed in \cite{resurrection} works in  $SU(5)\times U(1)_H$ 
grand unified models, with appropriate choices of the 
$U(1)_H$ charges such that the right-handed neutrinos are not very 
hierarchical, and all of them heavier than  $10^{12}~ GeV$.

 In models with grand unified gauge group 
$SO(10)$ and Pati-Salam, the mechanism does not work, since 
$\Delta L = 1$ right-handed neutrino 
decays are in equilibrium below $10^{12}~ GeV$, together with 
the $B+L$ violating sphaleron interactions.
 This is because in those models the Dirac neutrino mass matrix is 
somehow related to the up quark mass matrix, and thus it is 
very hierarchical. Since the observed hierarchy in the light 
neutrino masses is not so strong, 
$\sqrt{\Delta m^2_{atm}}/ \sqrt{\Delta m^2_{sol}} \sim 0.1$, 
this implies a hierarchy on the right-handed neutrinos,
 so one ore more of them result with a small mass, lighter than 
$10^{12}~ GeV$, and their lepton number violating decays are  
in thermal equilibrium. 
Although we have only analysed some particular models, this feature 
seems generic for such gauge groups, independently of the flavour 
symmetry, as far as neutrino masses are generated via the minimal 
seesaw mechanism, i.e., neglecting $SU(2)$ triplet vevs.

From a more phenomenological perspective, we have also found 
that GUT baryogensis works for many choices of the $U(1)_H$ 
charges in models with single right-handed neutrino dominance, and in models 
where two right-handed neutrinos dominate an lead to an inverted hierarchy 
of the left-handed neutrinos. Again, these models allow all heavy 
right-handed neutrinos with masses larger than $10^{12}~ GeV$.

\vskip 0.15in

{\bf Acknowledgements}
\vskip 0.15in
We are specially grateful to Sacha Davidson for many interesting 
comments and a careful reading of the manuscript.
We also thank Martin Hirsch for discussions, and all the organisers for 
a very interesting workshop.  This work was partially
supported by the Spanish MCyT grants BFM2002-00345 and
FPA2001-3031, by the TMR network contract HPRN-CT-2000-00148 of the
European Union and by the Generalitat Valenciana grant CTIDIB/2002/24.



\begin{thebibliography}{99}

\bibitem{gut-baryo}
A.D. Sakharov, {\sl Pis'ma\ Zh.\ Eksp.\ Teor.\ Fiz. }{\bf 5}, 32 (1967)
English translate: [A.D. Sakharov, \jetp5{24}{1967}];
S. Weinberg, \prl{42}{850}{1979};
A.Y. Ignatev, N.V. Krasnikov, V.A. Kuzmin and A.N. Tavkhelidze, 
\pl{B 76}{436}{1978};
M. Yoshi\-mu\-ra, \prl{41}{281}{1978} [Erratum: {\bf 43}, 746 (1979)]

\bibitem{sph}
V.A. Kuzmin, V.A. Rubakov, M.E. Shaposhnikov, \pl{B 155}{36}{1985};
N.S. Manton, \pr{D 28}{2019}{1983};
F.R. Klinkhammer, N.S. Manton \pr{D 30}{2212}{1984};
E.W. Kolb, M.S. Turner, {\sl The early Universe} (Addison-Wesley, 1990)

\bibitem{spheq}
P. Arnold, L. McLerran, \pr{D 36}{581}{87}; {\bf D 37}, 1020 (1988);
A. Bochkarev, M.E. Shaposhnikov, {\sl Mod.\ Phys.\ Lett.\ }, 
{\bf A 2}, 417 (1987); {\bf A 2}, 921 (1987);
S.Yu. Khlebnikov, M.E. Shaposhnikov, \np{B 308}{885}{1988};
D. B\"odeker, G.D. Moore, K. Rummukainen, 
{\sl Phys.~ Rev. } {\bf D 61}, 056003 (2000)

\bibitem{resurrection}
M. Fukugita, T. Yanagida, {\sl Phys.\ Rev.\ Lett. }{\bf 89}, 131602 (2002) 
[arXiv:hep-ph/0203194]

\bibitem{fukuda}
Super-Kamiokande Collaboration (Y. Fukuda \etal), \prl{81}{1562}{1998} 
[arXiv:hep-ex/9807003]

\bibitem{k2k}
K2K Collaboration, (S.H. Ahn \etal) \prl{90}{041801}{2003} 
[arXiv:hep-ex/0212007]

\bibitem{sksolar}
Super-Kamiokande Collaboration (S. Fukuda et al.), \pl{B 539}{179}{2002} 
[arXiv:hep-ex/0205075]

\bibitem{sno}
SNO Collaboration (Q.R. Ahmad \etal) \prl{89}{011302}{2002} 
[arXiv:nucl-ex/0204009]

\bibitem{kamland}
KamLAND Collaboration (K. Eguchi \etal) \prl{90}{021802}{2003} 
[arXiv:hep-ex/0212021]


\bibitem{seesaw}
M. Gell-Mann, P. Ramond, and R. Slansky, 
{\sl Supergravity} (North Holland,1979)];
T. Ya\-na\-gi\-da, {\sl Prog.\ Theor.\ Phys.}, {\bf 64}, 1103 (1980)];
R.N. Mohapatra and G. Senjanovi$\acute{c}$, \prl{44}{912}{1980}

\bibitem{us}
S. Davidson, J.A. Lopez-Perez, N. Rius, in preparation.


\bibitem{chooz}
CHOOZ Collaboration (M. Apollonio \etal),
\pl{B 466}{415}{1999} [arXiv:hep-ex/9907037]


\bibitem{U1}
H. Fritzsh, \pl{B 70}{436}{1977};
C.D. Froggatt, H.B. Nilsen, \np{B147}{277}{1979};
J. Harvey, P. Ramond, D. Reiss, \pl{B 92}{309}{1980};
C. Wetterich, \np{B 261}{461}{1985};
P. Kaus, S. Meshkov, \mpl{A 3}{1251}{1988};
L. Ib\'a\~nez, G.G. Ross, \pl{B 332}{100}{1994};
P. Binetruy, P. Ramond, \pl{B 350}{49}{1995} [arXiv:hep-ph/9412385]

\bibitem{nanwei-barr}
D.V. Nanopoulos, S. Weinberg, \pr{D 20}{2484}{1979};
S.M. Barr, G. Segre, H.A. Weldon, \pr{D 20}{2494}{1979}

\bibitem{buchyan}
W. Buchm\"uller, T. Yanagida, \pl{445}{339-402}{1999};
W. Buchm\"uller, M. Pl\"umacher, {\sl Int.\ J.\ Mod.\ Phys.}, {\bf A 15}, 
5047-5086 (2000) [arXiv:hep-ph/0007176], and references therein.


\bibitem{vala}
J. Sato, T. Yanagida, {\sl Nucl.~Phys.~Proc.~Suppl. }{\bf 77}, 293 (1999) 
[arXiv:hep-ph/9809307]

\bibitem{AF}
G. Altarelli, F. Feruglio,
in {\sl Neutrino Mass}, Springer Tracts in Modern Physics,
and references therein. 




\bibitem{triplets}
B. Bajc, G. Senjanovic and F. Vissani,
[arXiv:hep-ph/0210207]
 y Hirsch ,Valle ...


\bibitem{albright-barr}
S.M. Barr, S. Raby, \prl{79}{4748}{1997};
C.H. Albright, S.M. Barr, \pr{D 58}{013002}{1998};
C.H. Albright, K.S. Babu, S.M. Barr, \prl{81}{1167}{1998};
C.H. Albright, S.M. Barr, \prl{85}{244}{2000}; \pr{D 62}{093008}{2000}; 
\pr{D 64}{073010}{2001};
C.H. Albright, talk presented at Neutrinos and Implications for Physics Beyond 
the Standard Model, Stony Brook, New York, 11-13 Oct 2002 
[arXiv:hep-ph/0212090]



\bibitem{branco}
G.C. Branco, R. Gonzalez Felipe, F.R. Joaquim, M.N. Rebelo,  
\np{B 640}{202}{2002} [arXiv:hep-ph/0202030]

\bibitem{patisalam}
J.C. Pati, A. Salam, \pr{D 10}{275}{1974}

\bibitem{kingps}
S.F. King, M. Oliveira, \pr{D 63}{095004}{2001} [arXiv:hep-ph/0009287];
T. Blazek, S.F. King, J.K. Parry,  {\sl JHEP } {\bf 0305} 016 (2003) 
[arXiv:hep-ph/0303192]
 
\bibitem{king}
S.F. King, \np{B 576}{85-105}{2000} [arXiv:hep-ph/9912492], and
references therein.

\bibitem{inverted}
S.F. King, N. Nimai Singh, \np{B 596}{81}{2001} [arXiv:hep-ph/0007243]


\end{thebibliography}
\end{document}